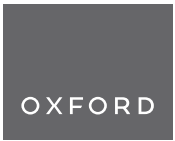

Sequence analysis

# iEnhancer-ELM: improve enhancer identification by extracting position-related multiscale contextual information based on enhancer language models

Jiahao Li[1,†], Zhourun Wu[1,†], Wenhao Lin[1], Jiawei Luo[1], Jun Zhang[1], Qingcai Chen [1,2] and Junjie Chen [1,*]

[1]School of Computer Science and Technology, Harbin Institute of Technology, Shenzhen, Guangdong 518055, China and [2]Guangdong Provincial Key Laboratory of Novel Security Intelligence Technologies, Harbin Institute of Technology, Shenzhen, Guangdong 518055, China

*To whom correspondence should be addressed.
†The authors wish it to be known that, in their opinion, the first two authors should be regarded as Joint First Authors.
Associate Editor: Guoqiang Yu

## Abstract

**Motivation:** Enhancers are important *cis*-regulatory elements that regulate a wide range of biological functions and enhance the transcription of target genes. Although many feature extraction methods have been proposed to improve the performance of enhancer identification, they cannot learn position-related multiscale contextual information from raw DNA sequences.

**Results:** In this article, we propose a novel enhancer identification method (iEnhancer-ELM) based on BERT-like enhancer language models. iEnhancer-ELM tokenizes DNA sequences with multi-scale $k$-mers and extracts contextual information of different scale $k$-mers related with their positions via an multi-head attention mechanism. We first evaluate the performance of different scale $k$-mers, then ensemble them to improve the performance of enhancer identification. The experimental results on two popular benchmark datasets show that our model outperforms state-of-the-art methods. We further illustrate the interpretability of iEnhancer-ELM. For a case study, we discover 30 enhancer motifs via a 3-mer-based model, where 12 of motifs are verified by STREME and JASPAR, demonstrating our model has a potential ability to unveil the biological mechanism of enhancer.

**Availability and implementation:** The models and associated code are available at https://github.com/chen-bioinfo/iEnhancer-ELM

**Contact:** junjiechen@hit.edu.cn

**Supplementary information:** Supplementary data are available at *Bioinformatics Advances* online.

## 1 Introduction

Enhancers are 50–1500 bp non-coding DNA fragments located at distal regions of target genes. They can *cis*-regulate target gene expression and control the process of transcription and translation of RNAs and proteins (Rong *et al.*, 2020). It has been found that enhancers are closely associated with several important disease genes, including oncogenes and tumor suppressor genes (Herz, 2016). Regarding the importance of enhancers in gene regularization, the identification of enhancers has become a research hot-spot, especially in the biological field of genome-wide enhancer annotation. Many computational methods have been developed for such issues.

Computational methods usually extract features from raw enhancer sequences firstly, and then train classifiers to predict enhancers. Thus, learning effective features is one of critical issues for improving the performance of enhancer identification. Depending on their feature extraction manners, computational methods can be categorized into three classes, including epigenetic information-based methods, sequence-component-based methods and deep learning-based methods. Epigenetic information-based methods transform external biological assessments, like genome-wide chromatin signatures (Firpi *et al.*, 2010) and histone epigenetic marks (Fernandez and Miranda-Saavedra, 2012), as enhancer features. However, they cannot be applied to cell lines with sparse data. Sequence-component-based methods extract features from enhancer sequences according to special rules for the inherent sequence properties. For example, iEnhancer-2L (Liu *et al.*, 2016) used the pseudo $k$-tuple nucleotide composition(PseKNC) (Chen *et al.*, 2014) to represent sequences;

iEnhancer-EL (Liu et al., 2018), iEnhancer-XG (Cai et al., 2021) and Enhancer-IF (Basith et al., 2021) integrated multiple rules to extract features from sequences. However, these special rules are insufficient as they are limited by human experience. Deep learning-based methods have achieved outstanding performance on enhancer identification by automatically learning powerful features in a data-driven manner. For example, iEnhancer-ECNN (Nguyen et al., 2019) and iEnhancer-GAN (Chen et al., 2021; Yang et al., 2021) used convolution neural networks(CNNs) for feature extraction. However, CNNs can only focus on local information (Tang et al., 2022). BiRen (Yang et al., 2017) used recurrent neural networks (RNNs) to extract features, but RNNs are vulnerable to gradient disappearance in the processing of long sequences.

In recent years, novel natural language processing (NLP) techniques (Ferruz and Höcker, 2022; Wu et al., 2023; Yan et al., 2023) (e.g. word embedding, attention mechanism and pre-trained language model) have shown the impressive ability to extract complex features from sentences. DNA is usually regarded as the 'language of life', whose alphabets are four nucleic acids (Li et al., 2021; Liu et al., 2019). Regarding this similarity between DNA sequences and natural language sentences, natural language techniques have been successfully applied in enhancer identification. EPIVAN (Hong et al., 2020) identified enhancers based on the nucleic acid embedding learned by dna2vec (Ng, 2017). iEnhancer-5step (Le et al., 2019) further improved the performance based on FastText (Bojanowski et al., 2017), which is an extension of dna2vec. However, due to the limitation of dna2vec that relies only on local information of neighbors, neither EPIVAN or iEnhancer-5step could capture globally contextual relation in enhancer sequences. To tackle such issues, pre-trained language models with attention mechanisms from NLP were introduced into enhancer identification. For example, BERT-Enhancer (Le et al., 2021) incorporated Bidirectional Encoder Representations from Transformers (BERT) (Devlin et al., 2019) as the embedding layer to convert raw enhancer sequences into vector representations. Although BERT-Enhancer achieved comparable performance with existing methods, it used a pre-trained BERT model on human language corpus. Therefore, the patterns captured by BERT-Enhancer lack interpretability. Currently, pre-trained biological language models, e.g. DNABERT (Ji et al., 2021) and TFBert (Luo et al., 2023) are the revolution in various computational biology tasks. However, the aim of these pre-trained models is to learn universal representations of biological sequences, but not for specific tasks. Thus, to facilitate the research on enhancer identification, it is urgent to construct a BERT-based enhancer language model. Currently, applying the language techniques to biological sequences remains challenging (Ferruz and Höcker, 2022), since words in human languages are natural, while words in biological sequences lack evidence. $k$-mers are usually regarded as the words of 'language of life'. However, the $k$-mers with different length contain multi-scale information (Jin et al., 2022; Yan et al., 2022). It is another challenge to integrate the multi-scale contextual information for enhancer identification.

In this article, we propose a novel enhancer identification method named iEnhancer-ELM, which tokenizes DNA sequences with different scale $k$-mers and captures the contextual information of $k$-mers by incorporating pre-trained BERT-based enhancer language models. The representations of DNA sequences are generated by averaging the tokens' embeddings projected by encoder layers. If the length of an input sequence exceeds the limitation of encoder layers, it will be split into several segments. The generated representations are then fed into an MLP layer for classification, as shown in Figure 1. We evaluate iEnhancer-ELM on two popular comprehensive benchmark datasets. Experimental results show the different scale $k$-mers have comparable performance on the same datasets, but their sensitivity and specificity differ greatly. By integrating multi-scale $k$-mers, iEnhancer-ELM outperforms existing state-of-the-art methods. To further illustrate the interpretability of attention mechanism, we show a case study that uses the model based on 3-mer to discovery motifs. We find 30 enriched enhancer motifs, where 12 of them are verified by a widely used motif discovery tool (STREME) and a popular transcription factor binding profiles dataset (JASPAR), demonstrating the ability of iEnhancer-ELM to unveil the biological mechanism of enhancer.

## 2 Methods

### 2.1 Benchmark datasets

To comprehensively evaluate iEnhancer-ELM, we acquire two popular benchmark datasets: Liu's dataset (Liu et al., 2016) and Basith's dataset (Basith et al., 2021). The sequences in Liu's dataset were selected from nine cell lines of humans. Moreover, all sequences were cropped to 200 bp and filtered by CD-HIT (Li and Godzik, 2006) to make sure their similarity less than 80%. For fair comparison, the benchmark dataset has been split into a training dataset and a test dataset. There are 1484 enhancers and 1484 non-enhancers in the training dataset, and 200 enhancers and 200 non-enhancers in the test dataset. As we would like to evaluate the capability of iEnhancer-ELM on enhancer identification and motif discovery, we only conduct experiments on the identification dataset (first phase), not on the strength dataset (second phase).

The Basith's dataset is a more comprehensive dataset, which contains eight subsets. The sequences in each subset are from one cell line, including epithelial (HEK293), normal human epidermal keratinocyte (NHEK), lymphoblast (K562), B lymphocyte (GM12878), human mammary epithelial cell (HMEC), human skeletal muscle myoblasts (HSSM), normal human lung fibroblasts (NHLF) and human umbilical vein endothelial cells (HUVEC). Different with the Liu's dataset in which the length of sequences is fixed, the length of sequences in Basith's dataset is variable ranging from 204 to 2000 bp. The redundancy was reduced to 60% by using CD-HIT for each cell line. Besides, this benchmark dataset has the same number between enhancer and non-enhancer sequences in the training sets, but in the test sets, the number of non-enhancer sequences is more than twice of enhancer sequences. Thus, Basith's dataset is more close to the real-world situations. The more detailed description of this dataset is shown in Basith et al. (2021).

### 2.2 Tokenization of enhancer sequences

An enhancer sequence in length of $L$ nucleotides can be represented as followed:

$$S = N_1 N_2 \ldots N_i \ldots N_L, \quad (1)$$

where $N_i$ denotes any of the four nucleotides $\{A, C, T, G\}$ at position $i$. We use overlapped $k$-mers as the 'words' of enhancer sequence, rather than a nucleotide, because $k$-mers contain richer contextual information. The overlapped $k$-mers are generated by a sliding window of $k$-size along the enhancer sequence. Therefore, the original sequence in length of $L$ nucleotides is tokenized into $L - k + 1$ $k$-mers. In this article, we choose $k = 3, 4, 5, 6$ considering the trade-off between vocabulary size and information in each token.

Besides, BERT-based language models have five special tokens, including classification token [CLS], padding token [PAD], unknown token [UNK], separation token [SEP] and mask token [MASK]. Thus, there are $4^k$ individual $k$-mers and five special tokens in the vocabulary.

### 2.3 Pre-trained BERT models on human genome corpus

We select a pre-trained BERT-based model, DNABERT (Ji et al., 2021) as an encoder to transform enhancer sequences to feature representations. DNABERT is composed of 12 attention layers and 12 attention heads in each layer. The attention mechanism is calculated by Eq. (2):

$$\text{MultiHead}(M) = \text{Concat}(\text{head}_1, \text{head}_2, \ldots, \text{head}_h) W^O$$

$$\text{head}_i = \text{softmax}\left(\frac{MW_i^Q (MW_i^K)^T}{\sqrt{d_k}}\right) \cdot MW_i^V \quad (2)$$

where $M$ is a hidden state of an input sequence, $h$ is the number of heads in an attention layer. $\{W_i^Q, W_i^K, W_i^V\}_{i=0}^h$ and $W^O$ are

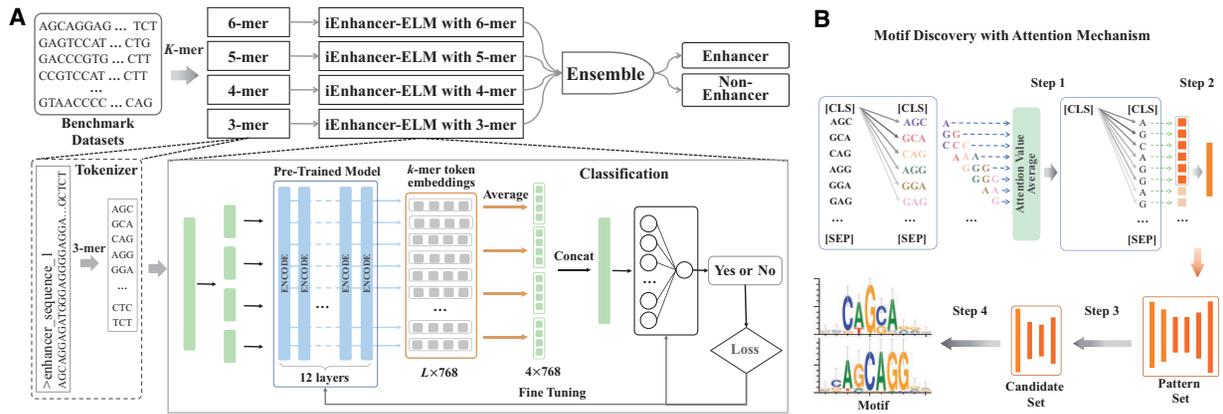

**Fig. 1.** Illustration of our proposed method. (**A**) The flowchart of enhancer identification. The input DNA sequences are first tokenized as different scale $k$-mers, which are then fed into corresponding enhancer language models. These models based on different scale $k$-mers are integrated together to identify enhancers. (**B**) Motif discovery via attention mechanism. There are four steps to find enriched enhancer motifs: calculate attention weight of each nucleotide from the special token [CLS], find potential patterns with high attention values, filter significant candidates by hypergeometric test and generate motifs according to sequence alignment

the attention matrices, and MultiHead() is the concatenate operation.

DNABERT learned the embeddings of $k$-mers via self-supervised learning on a masked language modeling (MLM) task, whose aim was to predict the masked tokens using contextual information. Different with the conventional mask mechanism that randomly masks partial tokens, DNABERT masked a span of $k$ tokens to avoid inferring one masked token directly from the overlapped $k$-mer tokens. The mask rate was set as 15%. The training corpus of DNABERT was generated by splitting a complete human genome (Harrow et al., 2012) into non-overlapping sub-sequences, where half of the sub-sequences were with length of 510 bp, and the other half were with random length between 10 and 509 bp.

### 2.4 Representation of enhancer sequences

We use DNABERT to learn the representation of enhancer sequences. However, DNABERT limits its input length to less than 510 tokens. As described above, we verify the validity of our model on two benchmark datasets. Liu's dataset consists of sequences in fixed length of 200 bps, but Basith's dataset consists of sequences in variable length with a range from 204 to 2000 bp.

For the Liu's dataset (Liu et al., 2016), all sequences are directly fed into DNABERT after tokenization. Each token ($k$-mer tokens and special tokens [CLS] and [SEP]) are projected into an embedding vector with 768 dimensions. We take the average of embedding vectors of all $k$-mer tokens as the representation of enhancer sequences. While For the Basith's dataset (Basith et al., 2021), due to the limitation of maximum input length of BERT-based models, we partition all sequences into 4 non-overlapping segments from left to right, and each segment comprises of 500 tokens, except the last segment that may have less than 500 tokens. These segments are then fed into DNABERT separately. We take the average of embedding vectors of each segment and then concatenate them to embedding vectors with length of 3072 as representations of enhancer sequences. Specially, for those enhancer sequences whose length is not enough to be divided into four segments, we will pad their final embedding vectors with 0 s until their length equals to 3072.

### 2.5 Fine tuning for enhancer identification

DNABERT was pre-trained on human genome data via self-supervised learning to capture the generally contextual relationship of nucleotide acids. Thus, it is not optimal for enhancer identification. We need to fine tune it to focus on the enhancers patterns.

We attach a 2-layer perceptron network as the classifier to the pre-trained model, and then fine-tune pre-trained weights on two enhancer benchmark datasets. The first layer is a hidden layer, and the second layer is an output layer to output a real value ranging in [0, 1] representing the probability of being an enhancer. A sequence is considered as an enhancer if its predicted probability is larger than a threshold. For training, following loss function is used:

$$\text{Loss} = \frac{1}{N}\sum_{i=0}^{N}[-y_i \log \hat{y}_i - (1-y_i)\log(1-\hat{y}_i)] + \lambda ||W||_2^2 \quad (3)$$

where the first term is the cross entropy loss, and the second term is the $L_2$ normalization.

### 2.6 Motif discovery via attention mechanism

We analyze the captured biological patterns by enhancer language models via exploring the weights in attention mechanism (Fig. 1B). Those biological patterns passed the hypergeometric test are then considered as motifs. Each enhancer language model contains 12 embedding layers and each layer contains 12 attention heads. Since the last layer has the highest distinguishing power to identify enhancers, we find the biological patterns of enhancers in the average of the 12 attention matrices of the last layer.

**Step 1: calculate the attention weight of a nucleotide in an enhancer sequence.** Since the vocabulary is composed of $k$-mer tokens and special tokens, the attention weights represent the relationship between tokens, rather than a nucleotide. We only consider the attention between [CLS] and $k$-mer tokens, as the [CLS] is usually used to represent the entire sequence. The attention of each $k$-mer from [CLS] is first equally distributed among its nucleotides, and then all attention of a nucleotide is averaged as its attention weight in an enhancer sequence. This attention weight represents the importance of the corresponding nucleotide to the entire original sequence.

**Step 2: find out the potential patterns.** The potential patterns are found out from above attention vectors by two criteria: (i) all attention weights of nucleotides in a pattern are larger than the average of the attention vector; (ii) the length of patterns is not less than five nucleotides with at most one gap. For facilitating the pairwise sequence alignment, we expand the length of a potential pattern region to 10 nucleotides along both directions of the original sequence.

**Step 3: filter significant candidates.** For each potential pattern, we perform a hypergeometric test (Beal, 1976) to calculate the P-value and determine whether the candidate is significant.

$$P_{\text{value}}(X) = 1 - \frac{\binom{K}{k}\binom{N-K}{n-k}}{\binom{N}{n}} \quad (4)$$

where $X$ represents a candidate pattern, $K$ is the number of positive sequences, $k$ is the number of regions $X$ occurring in positive sequences, N is the number of all sequences and n is the number of patterns X occurring in all sequences. Here, we apply the AhoCorasick algorithm (Aho and Corasick, 1975) for the computation of n and k. The significance threshold is set as 0.005.

**Step 4: generate motifs according to the sequence alignment.** We first find motif candidates with high sequence similarity by iterative pairwise alignment, then calculate the nucleotide frequency at each position to obtain motifs. In each iteration, the longest or the most frequent sequence will be selected as a core to align with remaining candidates one by one. If the matched positions exceed the half length of the alignment, this region will be categorized into the same group of the core. At last, we remove groups that have less than 11 members, and then calculate a position-weight matrix of the remaining groups for producing motifs.

### 2.7 Performance metrics

In this study, we select six metrics to evaluate the performance of enhancer prediction: sensitivity (Sn), specificity (Sp), accuracy (Acc), Matthews correlation coefficient (MCC), balanced accuracy (Bacc) (Basith et al., 2021) and Area Under ROC Curve[AUC (Fawcett, 2006)]. These metrics are defined as:

$$Sn = \frac{TP}{TP+FN}; \quad Sp = \frac{TN}{TN+FP}$$
$$Acc = \frac{TP+TN}{TP+TN+FP+FN}; \quad Bacc = \frac{Sn+Sp}{2}$$
$$MCC = \frac{(TP \times TN) - (FP \times FN)}{\sqrt{(TP+FP)(TP+FN)(TN+FP)(TN+FN)}}$$

where TP, TN, FP and FN represent true positive, true negative, false positive and false negative, respectively. Sn is the proportion of actual positive cases that have been predicted as positive, and Sp is the proportion of actual negatives. Acc is the proportion of all samples that their predictions are consistent with real labels. Compared with Acc, Bacc is more suitable to evaluate performance on an imbalanced dataset. The Matthews correlation coefficient (MCC) is a more reliable statistical rate which produces a high score only if the prediction obtained good results in all of the four confusion matrix categories. Specifically, the range of MCC values is [−1, 1]. AUC is a comprehensive metric, because it provides an aggregate measure of performance across all possible thresholds. AUC ranges in value from 0 to 1.

## 3 Results

### 3.1 Experimental setup

In this study, all DNA sequences are tokenized with k-mers (k = 3, 4, 5, 6). And all enhancer language models have 12 attention layers as encoder layers and 2 MLP layers as classification layers. Each attention layer has 12 attention heads. The embedding dimension of each token is 768. After being extracted by encoder layers, features are fed into two 2 MLP layers with one hidden layer of 25 neurons for Liu's dataset or 128 neurons for Basith's dataset. To preserve previous knowledge and avoid catastrophic forgetting, we fine-tune the pre-trained model with a small learning rate by exponential decay. In addition, we add a dropout layer with a drop rate of 0.3 after the hidden layer to prevent overfitting. We use Adam as the optimizer. iEnhancer-ELM is implemented with PyTorch 1.9.0 equipped with NVIDIA A100 80GB.

However, due the different characters of these two benchmark datasets, we use different training tricks for them. For the Liu's benchmark dataset, all models are fine-tuned for 30 epochs. Besides, we regard the threshold as a hyperparameter and find the threshold of 0.55 is the best trade-off between sensitivity and specificity. For the Basith's benchmark dataset, the original training dataset of each cell line is divided into validation dataset and training dataset with a ratio of 1:9. The final models on eight cell lines are selected according to their performance in validation dataset. In order to obtain stable performance on all cell lines, we use the adversarial training (Miyato et al., 2017) and batch normalization during the training process. More details of experiment configuration are shown in GitHub (https://github.com/chen-bioinfo/iEnhancer-ELM.git).

### 3.2 The performance of enhancer language models with different scale k-mers

In this article, all DNA sequences are tokenized with different k-mers (k = 3, 4, 5, 6) and then fed into enhancer language models. Thus, we first evaluate the performance of different scale k-mers.

Table 1 shows the results of enhancer language models with different scale k-mers on Liu's test dataset. Different scale k-mers achieve comparable performance. Specifically, 3-mer achieves the best performance with an AUC of 0.8485, followed by 4-mer, 5-mer and 6-mer, respectively. While 3-mer and 5-mer achieve the same performance in terms of Acc, followed by 4-mer and 6-mer, respectively. We also compare their performance with several widely-used features, including manual features [e.g. naive k-mer (Lee et al., 2011), PseDNC (Chen et al., 2013), spectrum profile (Liu et al., 2015) and mismatch profile (Liu et al., 2015)] and data-driven features [e.g. FastText (Le et al., 2019), Word2Vec (Hong et al., 2020) and BERT-Enhancer (Le et al., 2021)]. The comparison is conducted with the 5-fold cross validation on the Liu's training dataset, and the results are shown in Supplementary Table S1. We find that the different scale k-mers learned by enhancer language models outperform those widely-used features.

Table 2 shows the performance of enhancer language models with different scale k-mers on eight cell lines in the Basith's dataset. For each cell line, the different scale k-mers achieve comparable performance in terms of AUC. 3-mer achieves the best performance on five cell lines, including HEK293, K652, GM12878, HMEC and HSMM. 4-mer achieves the best performance on two cell lines, including NHLF and HUVEC. 6-mer achieves the best performance on NHEK, respectively. But their performance on different cell lines differs largely, such as 3-mer achieves an AUC of more than 0.9 on HEK293 and GM12878, while it only achieves an AUC of about 0.78 on NHEK and HUVEC.

Besides, to investigate the effect of fine-tuning, we conduct additional experiments that freeze the weights of pre-trained encoder layers and only update the weights of classification layers. The results are shown in Supplementary Tables S2 and S4. We find that the fine-tuned enhancer language models outperform the pre-trained methods on the Liu's dataset. But on Basith's dataset, they achieve comparable performance. To further investigate the nuanced effect of fine-tuning, we also visualize the distribution of enhancers and non-enhancers projected by all encoder layers on the Liu's training dataset (Supplementary Fig. S1). It is obvious that the distributions of enhancers and non-enhancers projected by pre-trained models are mixed together. While the distributions projected by fine-tuned models are divided into two clusters, which is more obvious at higher encoder layers. These results indicate that fine-tuning can make the pre-trained model focus more on task-specific information. Thus, all following experiments are conducted via fine-tuning manner.

### 3.3 The different scale k-mers are complementary

Although different scale k-mers have comparable performance on most datasets, they still have different performance in terms of sensitivity and specificity. Thus, we then evaluate the complement of different scale k-mers. The experiments are conducted from two views:

**Table 1.** The performance of fine-tuned enhancer language models with different scale k-mers on Liu's test dataset

| k-mers | Acc | Sn | Sp | MCC | AUC |
|---|---|---|---|---|---|
| 3-mer | 0.8075 | 0.7900 | 0.8250 | 0.6154 | 0.8485 |
| 4-mer | 0.7975 | 0.7550 | 0.8400 | 0.5972 | 0.8331 |
| 5-mer | 0.8075 | 0.7900 | 0.8250 | 0.6154 | 0.8326 |
| 6-mer | 0.7950 | 0.7650 | 0.8250 | 0.5911 | 0.8116 |

Table 2. The performance of fine-tuned enhancer language models with different scale *k*-mers on Basith's test dataset

| Cell line | *k*-mers | Bacc | Sn | Sp | MCC | AUC |
|---|---|---|---|---|---|---|
| HEK293 | 3-mer | **0.8608** | 0.8479 | 0.8736 | **0.7057** | **0.9346** |
|  | 4-mer | 0.8520 | 0.8280 | **0.8759** | 0.6917 | 0.9323 |
|  | 5-mer | 0.8278 | 0.8509 | 0.8561 | 0.6868 | 0.9264 |
|  | 6-mer | 0.8278 | **0.8798** | 0.8412 | 0.7033 | 0.9289 |
| NHEK | 3-mer | 0.7095 | 0.7877 | 0.6312 | 0.3950 | 0.7821 |
|  | 4-mer | 0.7204 | 0.7447 | 0.6960 | 0.4175 | 0.7924 |
|  | 5-mer | 0.7052 | **0.7980** | 0.6123 | 0.3872 | 0.7614 |
|  | 6-mer | **0.7766** | 0.7229 | **0.8303** | **0.5453** | **0.8716** |
| K652 | 3-mer | **0.7802** | 0.7840 | 0.7764 | **0.5371** | **0.8625** |
|  | 4-mer | 0.7777 | 0.7672 | **0.7882** | 0.5351 | 0.8589 |
|  | 5-mer | 0.7759 | **0.7862** | 0.7655 | 0.5273 | 0.8531 |
|  | 6-mer | 0.7609 | 0.7857 | 0.7541 | 0.5146 | 0.8488 |
| GM12878 | 3-mer | **0.8192** | **0.8022** | 0.8362 | 0.6211 | **0.9121** |
|  | 4-mer | 0.8137 | 0.7555 | 0.8718 | 0.6255 | 0.9083 |
|  | 5-mer | 0.8073 | 0.7946 | 0.8200 | 0.5955 | 0.9047 |
|  | 6-mer | 0.8163 | 0.7445 | **0.8880** | **0.6376** | 0.8907 |
| HMEC | 3-mer | **0.7645** | 0.7638 | **0.7652** | **0.5068** | **0.8631** |
|  | 4-mer | 0.7585 | 0.7560 | 0.7610 | 0.4953 | 0.8282 |
|  | 5-mer | 0.7608 | **0.7911** | 0.7304 | 0.4948 | 0.8292 |
|  | 6-mer | 0.7399 | 0.7371 | 0.7426 | 0.4586 | 0.8119 |
| HSMM | 3-mer | 0.7193 | 0.7191 | **0.7194** | 0.4179 | **0.7948** |
|  | 4-mer | **0.7218** | **0.7520** | 0.6915 | **0.4197** | 0.7876 |
|  | 5-mer | 0.7031 | 0.7355 | 0.6707 | 0.3840 | 0.7702 |
|  | 6-mer | 0.7012 | 0.6967 | 0.7056 | 0.3832 | 0.7712 |
| NHLF | 3-mer | **0.7801** | 0.8058 | 0.7544 | **0.5331** | 0.8501 |
|  | 4-mer | 0.7754 | 0.7767 | **0.7741** | 0.5280 | **0.8544** |
|  | 5-mer | 0.7545 | 0.7665 | 0.7424 | 0.4849 | 0.8244 |
|  | 6-mer | 0.7655 | **0.8325** | 0.7005 | 0.5030 | 0.8377 |
| HUVEC | 3-mer | **0.7235** | 0.7487 | 0.6983 | 0.4236 | 0.7880 |
|  | 4-mer | 0.7224 | **0.7663** | 0.6786 | 0.4202 | **0.7979** |
|  | 5-mer | 0.7248 | 0.7534 | 0.6962 | **0.4257** | 0.7897 |
|  | 6-mer | 0.7190 | 0.7284 | **0.7095** | 0.4162 | 0.7923 |

The bold entries is to highlight the best results under equivalent experiments.

Table 3. Performance comparison between iEnhancer-ELM and the state-of-the-art methods on the Liu's dataset

| Method | Acc | Sn | Sp | MCC | AUC |
|---|---|---|---|---|---|
| iEnhancer-2L | 0.7300 | 0.7100 | 0.7500 | 0.4600 | 0.8062 |
| iEnhancer-EL | 0.7475 | 0.7100 | 0.7850 | 0.4960 | 0.8173 |
| iEnhancer-XG | 0.7575 | 0.7400 | 0.7750 | 0.5140 | – |
| iEnhancer-Deep | 0.7402 | 0.8150 | 0.6700 | 0.4902 | – |
| BERT-Enhancer | 0.7560 | 0.8000 | 0.7750 | 0.5150 | – |
| iEnhancer-GAN | 0.7840 | 0.8110 | 0.7580 | 0.5670 | – |
| iEnhancer-5Step | 0.7900 | **0.8200** | 0.7600 | 0.5800 | – |
| iEnhancer-ECNN | 0.7690 | 0.7850 | 0.7520 | 0.5370 | 0.8320 |
| iEnhancer-ELM | **0.8300** | 0.8000 | **0.8600** | **0.6612** | **0.8560** |

The bold entries is to highlight the best results under equivalent experiments.

Thirteen samples are predicted correctly by only one model, such as five samples are predicted correctly by only 3-mer, three samples by 4-mer, three samples by 5-mer and two samples by 6-mer. Forty-nine samples are predicted correctly by two or three models. These results demonstrate the different scale *k*-mers learned by enhancer language models are complementary.

We have the same conclusion on the Basith's dataset. The complementary analysis for 8 cell lines in Basith's dataset is shown in Supplementary Figure S2. The distribution of different scale *k*-mers in all cell lines does not overlap. For the intersection of prediction results, only 72.1% of the samples (including enhancers and non-enhancers) on average in eight cell lines are accurately predicted by all models. And 8.21% of samples on average were predicted correctly by one model, and 7.98% by two models and 11.71% by three models.

### 3.4 iEnhancer-ELM outperforms competing methods

According to above observations, different scale *k*-mers are complementary for enhancer identification. Thus, we integrate multi-scale *k*-mers as one model iEnhancer-ELM for improving the performance on enhancer identification and compete it with several state-of-the-art methods.

Table 3 shows the comparison between iEnhancer-ELM and existing state-of-the-art methods on the Liu's dataset. The competing methods include iEnhancer-2L (Liu et al., 2016), iEnhancer-EL (Liu et al., 2018), iEnhancer-5Step (Le et al., 2019), BERT-Enhancer (Le et al., 2021), iEnhancer-XG (Cai et al., 2021), iEnhancer-GAN (Yang et al., 2021), iEnhancer-ECNN (Nguyen et al., 2019) and iEnhancer-Deep (Kamran et al., 2022). iEnhance-2L, iEnhancer-EL and iEnhancer-XG extract the inherent properties from nucleic acid sequences based on manual rules. iEnhancer-Deep, iEnhancer-GAN and iEnhancer-ECNN represent raw enhancer sequences with one-hot encoding and combine CNN to extract features for classification. iEnhancer-5Step and BERT-Enhancer use pre-trained FastText and BERT to learn representations of raw sequences, respectively. None of competing method achieves an accuracy of larger than 0.8000, while iEnhancer-ELM achieves an accuracy of 0.8300 and an AUC of 0.8560, outperforming existing state-of-the-art methods. Table 4 shows the comparison between iEnhancer-ELM and Enhancer-IF (Basith et al., 2021) on the eight cell lines in Basith's dataset. iEnhancer-ELM outperforms Enhancer-IF on six cell lines and achieves comparable performance on the other two cell lines in terms of AUC. Enhancer-IF is an integrated method based on five machine learning methods (e.g. random forest, extremely randomized tree, multilayer perceptron, support vector machine and extreme gradient boosting) and seven types of features (e.g. Kmer, CKSNAP, EIIP, DPCP, TPCP, PseDNC, PseTNC) extracted by manual rules. However, iEnhancer-ELM just uses an enhancer language model based on multi-scale *k*-mers. Thus, iEnhancer-ELM is not only more distinguishable but also more convenient to implement than Enhancer-IF.

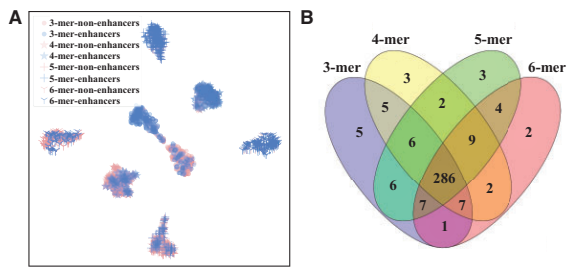

Fig. 2. Complementary analysis of different scale *k*-mers on the Liu's dataset. (**A**) t-SNE visualization of distribution of enhancer and non-enhancer sequences. (**B**) Venn diagram of prediction results

the distribution of enhancers and non-enhancers based on different scale *k*-mers and the intersection of predicted results.

Figure 2 shows the complementary analysis of different scale *k*-mers on the Liu's dataset. We are surprised to find that the distribution of all *k*-mers is centrosymmetric. The enhancers (blue color) lie in the upper right of the space, and the non-enhancers (pink color) are in the lower left corner of the space. Specifically, the 3-mer is distributed in the center, the enhancers and non-enhancers encoded by 4-mer, 5-mer and 6-mer are partitioned into two sides. For the intersection of prediction results, there are 200 enhancers and 200 non-enhancers in total, where only 286 samples (including enhancers and non-enhancers) are predicted correctly by all models.

Table 4. Performance comparison between iEnhancer-ELM and the state-of-the-art methods on the eight cell lines on the Basith's dataset

| Cell line | Method | Bacc | Sn | Sp | MCC | AUC |
|---|---|---|---|---|---|---|
| HEK293 | Enhancer-IF | 0.8170 | 0.7750 | 0.8580 | 0.6250 | 0.8930 |
|  | iEnhancer-ELM | 0.8732 | 0.8666 | 0.8798 | 0.7283 | 0.9443 |
| NHEK | Enhancer-IF | 0.7480 | 0.7340 | 0.7620 | 0.4770 | 0.8100 |
|  | iEnhancer-ELM | 0.7766 | 0.7229 | 0.8303 | 0.5453 | 0.8716 |
| K652 | Enhancer-IF | 0.7730 | 0.7730 | 0.7720 | 0.5230 | 0.8520 |
|  | iEnhancer-ELM | 0.7974 | 0.8180 | 0.7767 | 0.5679 | 0.8712 |
| GM12878 | Enhancer-IF | 0.8140 | 0.7550 | 0.8730 | 0.6260 | 0.9010 |
|  | iEnhancer-ELM | 0.8222 | 0.7564 | 0.8879 | 0.6475 | 0.9179 |
| HMEC | Enhancer-IF | 0.7530 | 0.7400 | 0.7650 | 0.4850 | 0.8390 |
|  | iEnhancer-ELM | 0.7645 | 0.7638 | 0.7652 | 0.5068 | 0.8631 |
| HSMM | Enhancer-IF | 0.7180 | 0.7090 | 0.7270 | 0.4170 | 0.7940 |
|  | iEnhancer-ELM | 0.7193 | 0.7191 | 0.7194 | 0.4179 | 0.7948 |
| NHLF | Enhancer-IF | 0.7940 | 0.7890 | 0.7980 | 0.5650 | 0.8650 |
|  | iEnhancer-ELM | 0.7884 | 0.8236 | 0.7532 | 0.5479 | 0.8623 |
| HUVEC | Enhancer-IF | 0.7370 | 0.7520 | 0.7220 | 0.4500 | 0.8110 |
|  | iEnhancer-ELM | 0.7334 | 0.7691 | 0.6977 | 0.4417 | 0.8045 |

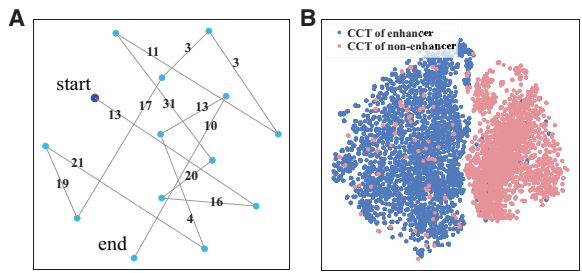

**Fig. 3.** The distribution of embeddings of *CCT*s learned by iEnhancer-ELM on the Liu's dataset. (**A**) The distribution of all *CCT*s in the enhancer sequence Chr19_39177160_39177360. The navy blue dot is the first *CCT*, followed by other *CCT*s (light blue dots). The numbers in lines mean their nucleotide distance in the sequence. (**B**) The distribution of embeddings of *CCT*s in all enhancers and non-enhancers

### 3.5 Enhancer language models capture position-related contextual information of *k*-mers

Most competing methods also take the *k*-mers as tokens to encode raw sequences, especially the naive *k*-mers (Lee *et al.*, 2011) and dna2vec (Hong *et al.*, 2020). However, iEnhancer-ELM outperforms these competing methods, which is because iEnhancer-ELM captures the position-related contextual information of *k*-mers. The impact of *k*-mers on enhancer identification usually depends not only on their frequencies but also on their positions.

We present a case study to illustrate that contextual information of *k*-mers learned by enhancer language models changes according to their positions in DNA sequences. Figure 3 shows the distribution of embeddings of *CCT*s on the Liu's dataset. We first visualize the distribution of all *CCT*s in the enhancer sequence Chr19_39177160_39177360, which contains 14 non-overlapped *CCT*s. The embeddings of *CCT*s change according to their positions in the raw sequence (Fig. 3A). In general, these *CCT*s that are close in sequence positions are also close in the embedding space. Figure 3B shows the distribution of *CCT*s in all enhancers and non-enhancers. We observe a clear gap between the *CCT*s from enhancer sequences and those from non-enhancer sequences. These results demonstrate that position-related contextual information of *k*-mers have a powerful ability to discriminate enhancers.

### 3.6 Enhancer language models can be used to discovery motifs

Since enhancer language models pay more attention to patterns that have important contributions to enhancer identification, these important patterns can be extracted by large attention weights in encoder layers. Those patterns passed the hypergeometric test are considered as motifs. We use the enhancer language model based on 3-mer fine-tuned on the Liu's dataset as an instance to illustrate the motif discovery. We discover 30 motifs as shown in Figure 4. The left panel shows two motifs that have high attention weights in corresponding regions, while the right panel shows all discovered motifs. To illustrate the reliability of these motifs, we compare our discovered motifs with those motifs found by a widely-used motif discovery tool STREME (Bailey, 2021) and a popular transcription factor binding database JASPAR (Castro-Mondragon *et al.*, 2022) via TomTom (Gupta *et al.*, 2007), as shown in Figure 5.

**Comparison to STREME for finding enriched motifs.** STREME (Bailey, 2021) is an effective motif discovery tool to find ungapped motifs that are enriched in one dataset compared to a control dataset. We conduct STREME on the Liu's dataset and find 8 motifs, as shown in Supplementary Figure S3, in which 6 motifs are matched with 10 motifs found by enhancer language models. The comparison is shown in Supplementary Figure S4. Note that several motifs may be matched with the same one.

**Comparison to existing transcription factor binding sites in JASPAR database.** JASPAR is an open-access database storing manually curated transcription factor (TF) binding profiles as position frequency matrices (PFMs) (Castro-Mondragon *et al.*, 2022). TFs bind to either enhancer or promoter regions of DNA adjacent to the genes that they regulate. Thus, transcription factor binding sites (TFBS) are vital for enhancer function. In practice, we compare the discovered motifs with TF binding profiles in the JASPAR database. There are six motifs significantly matched with TFBS in JASPAR (Supplementary Fig. S5).

These results show that the attention mechanism in enhancer language models has an ability to discover the enriched motifs and functional regions in enhancer sequences.

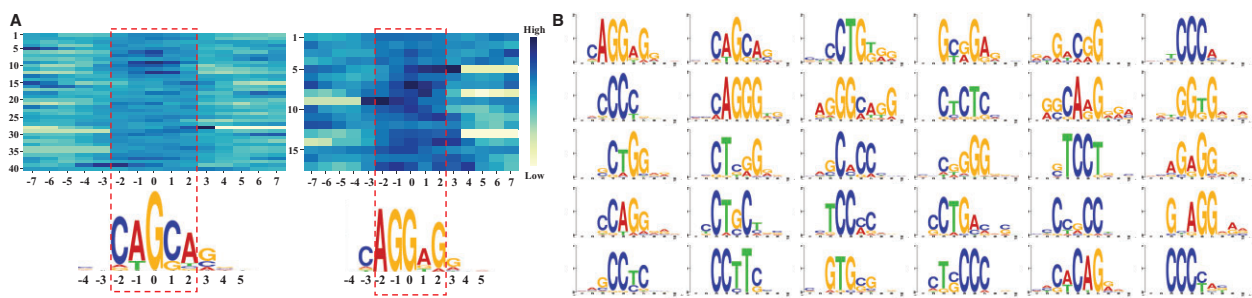

**Fig. 4.** Motif discovery via attention mechanism in enhancer language models. (**A**) Motifs have high attention weights in corresponding regions. In the heatmap, each row represents a segment that passes the hypergeometric test, and each column represents a nucleotide. The *x*-axis is centered on the core regions of biological patterns, and the *y*-axis is the number of segments in the sequence alignment. The highlighted regions are the motifs captured by attention mechanism. (**B**) Thirty discovered motifs by the enhancer language model based on 3-mer on the Liu's dataset

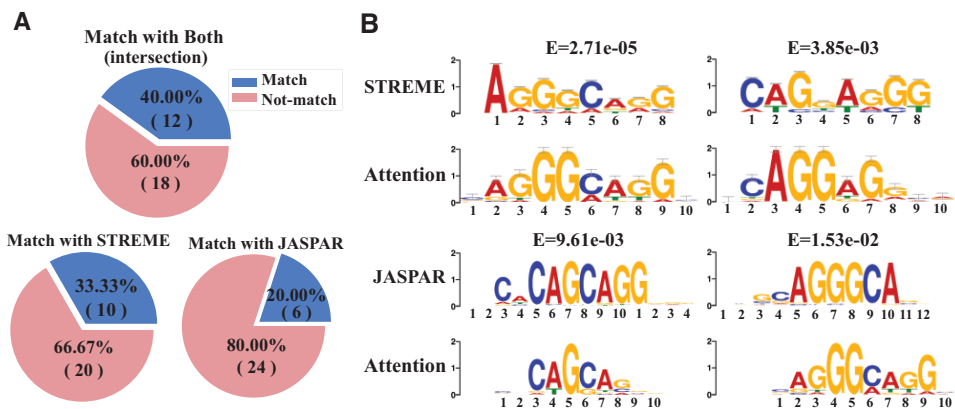

**Fig. 5.** Verification of discovered motifs. (**A**) The percentage of discovered motifs that are matched with STREME and JASPAR. (**B**) Four example motifs that are significantly matched

## 4 Conclusion

In this study, we propose a novel enhancer identification method named iEnhancer-ELM based on BERT-like enhancer language models. iEnhancer-ELM captures position-related contextual information of multi-scale $k$-mers to improve the performance of enhancer identification. The results on two benchmark datasets show that iEnhancer-ELM outperforms state-of-the-art methods. Moreover, we illustrate the interpretability of enhancer language models by showing a case study that captures 30 motifs, where 12 of discovered motifs can be matched with widely used motif discovery tools STREME and JASPAR, demonstrating our model has a potential ability to reveal the biological mechanism. For future works, we will explore the enhancer identification across tissues or species.

## Author contributions

J.C. conceptualized the study and designed the experiment. J.L. collected the enhancer data. J.L., Z.W. and W.L. implemented the algorithms. J.L., Z.W., J.L. and J.Z. performed the analysis. J.C., J.L. and Q.C. wrote and reviewed the manuscript.

## Funding

This work was supported by the Natural Science Foundation of China [62102118], Project of Educational Commission of Guangdong Province of China [2021KQNCX274], Guangdong Provincial Key Laboratory of Novel Security Intelligence Technologies [2022B1212010005] and the Shenzhen Colleges and Universities Stable Support Program [GXWD20220811170504001].

*Conflict of Interest*: none declared.